\documentclass[aps,prb,twocolumn]{revtex4}
\usepackage{amssymb}
\usepackage{epsfig}
\usepackage{float}
\usepackage{graphicx}
\usepackage{epsfig}
\usepackage{float}
\usepackage{amsmath}

\begin{document}

\title{Absence of magnetic ordering in NiGa$_{2}$S$_{4}$}
\author{I.I. Mazin}
\affiliation{Code 6390, Naval Reserach Laboratory, Washington, DC 20375}
\begin{abstract}
Triangular-layered NiGa$_{2}$S$_{4}$, contrary to intuitive expectation,
does not form a noncollinear antiferromagnetic structure, as do isoelectronic
NaCrO$_{2}$ and LiCrO$_{2}$. Instead, the local magnetic moments remain
disordered down to the lowest measured temperature. To get more insight into
this phenomenon, we have performed first principles calculations of the first, second 
end third neighbors exchange interactions, and found that the second
neighbor exchange is negligible, while the 
first and the third 
neighbor exchanges are comparable and antiferromagnetic. Both are 
rapidly suppressed by the on-site Hubbard repulsion.
\end{abstract}
\date{\today }
\maketitle

NiGa$_{2}$S$_{4}$ occurs in the layered structure where the main motif is a
triangular layer of Ni$^{2+}$ ions surrounded by edge-sharing S octahedra,
forming a trilayer S-Ni-S with the rhombohedral stacking ABC, and the trilayers
are separated by the gallium oxide layers. Ni$^{2+}$ has the electronic
configuration $t_{2g}^{6}e_{g}^{2},$ and one expects it to be insulating and
magnetic with $S=1,$ and the magnetic moment per Ni being somewhat less than 
$2$ $\mu _{B}.$ Indeed, this is exactly what happens with structurally
similar transition metal oxides with a transition metal in the $d^{8}$
configuration, such as NaCrO$_{2}$ or LiCrO$_{2}.$ 3$d$ metal ions in this
confuration do not have an orbital moment, therefore one expects a vanishing
single-site anisotropy and magnetic interactions reasonably well described
by the Heisenberg model. In the nearest neighbor approximation this leads to
noncollinear ground states, with neighboring spins pointing roughly at 120$%
^{o}$ to each other. Indeed, this is what has been observed in the
above-mentioned chromates.

NiGa$_{2}$S$_{4},$ on the other hand, has attracted substantial recent
interest exactly because the experiments indicate absense of any long-range
magnetic ordering\cite{Science}. Several explanations have been proposed,
such as full cancellation of the nearest neighbor exchange and frustrated
competition between the 2nd and the 3rd neighbor interactions\cite{Science},
or biquadratic exchange\cite{biq}. These, however, impose very severe
quantitative restriction on the exchange parameters, which seem
quite unrealistic.

In order to elucidate magnetic interactions in this system we have performed
ab initio density functional theory (DFT) calculations of the electronic
structure and magnetic energies of NiGa$_{2}$S$_{4}$ and found that indeed
condition required by either explanation are very unlikely to be satisfied,
however, the magnetic interactions are very rich and (as
conjectured in Ref. \cite{Science}) long
range, so that taking into account  three neighbor shells is
indispensable.

\begin{figure}[htbp]
\centering
\includegraphics[angle=0,width=0.99\linewidth]{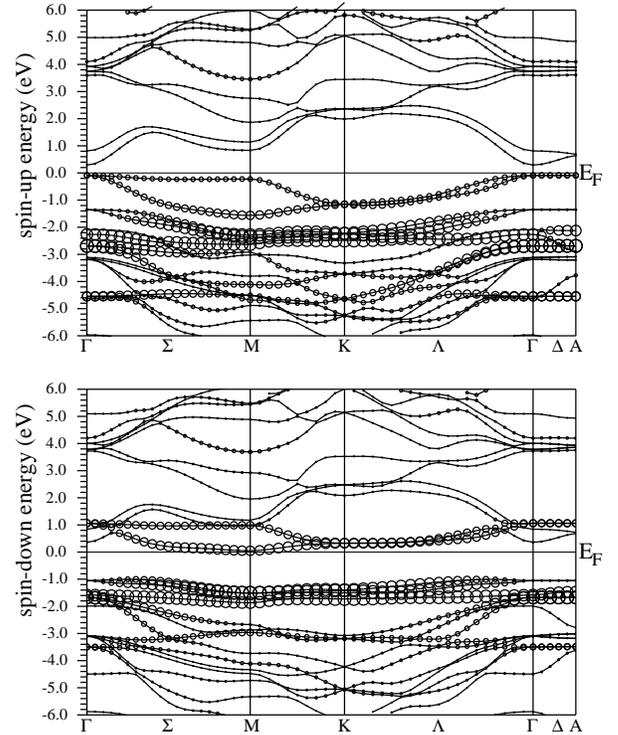}
\caption{LDA band structure of the ferromagnetic NiGa$_2$O$_4$. The Ni character
is emphasized by the size of the circles.}
\label{bands}
\end{figure}

For the calculations, the experimental crystal structure\cite{str} was used.
A full potential linear augmented plane wave code\cite{Wien2k} was used with
a gradient approximation for exchange and correlation\cite{PBE}. The
calculations, as expected, render an insulating band structure shown in Fig.
1. As one can see, Ni $d(e_{g})$ bands are fully polarized, a small gap
opens (as usual, the absolute value of the gap in the DFT cannot be taken
very seriously), the magnetic moment per Ni is 2 $\mu _{B},$ and in the
calculations this moment, not unexpectedly, resides entirely in the NiS$_{2}$
layer. Interestingly, more than 20\% of the total magnetic moment is located
on the S sites. This creates a ferromagnetic interaction between the nearest
neighbor nickels, which in the theory of strongly correlated magnetic
systems is known as \textquotedblleft ferromagneting 90$^{o}$
exchange\textquotedblright . Note that in the DFT the Hund rule energy is
approximated as $\int d\mathbf{r}I(\mathbf{r)}m^{2}(\mathbf{r})/4\approx
\sum I_{i}M_{i}^{2}/4,$ where $m(\mathbf{r)}$ is the total spin density, $I$
and $M_{i}$ are the Stoner factor and the total magnetization of the atom $%
i, $ and this energy therefore does not depend on the Ni-S-Ni bond angle.
Let us estimate this interaction\cite{note1}. Two Ni neighbors, when their
spins are parallel, induce a magnetic moment of $\sim 0.2$ $\mu _{B}$ on
each of the two bridging sulfurs, gaining an additional magnetic energy of 2$%
I_{S}0.2^{2}/4\approx 20$ meV (The Stoner factor of the sulfur ion can be
estimated as described in Ref.\onlinecite{Mol} and is about 1 eV). Defining $J$ as
half of the energy for flipping a bond ($E_{nn^{\prime }}=J_{nn^{\prime }}%
\mathbf{S}_{n}\cdot \mathbf{S}_{n^{\prime }}),$ we find a ferromagnetic
contribution to $J$ of the order of 10 meV. This is not a small energy, but it
has to compete with also large conventional antiferromagnetic superexchange.
The latter in DFT is 2$t^{2}/I_{Ni},$ where $t$ is the effective Ni-Ni $d-d$
hopping and $I_{Ni}$ is the Hund energy cost of exciting an electron
with the opposite spin, $%
I_{N}\approx 0.8$ eV. However, taking into account on-site Mott-Habbard
correlation mandates substiting the Stoner $I$ in this expression by the
Hubbard $U$, which is at least 4 times larger. While Ni in NiGa$_{2}$S$_{4}$
is not necessarily strongly correlated, there is no doubt that the energy of
adding an electron 
is substantially underestimated in the DFT, and hence the AFM
superexchange overestimated. 
\begin{figure}[tbph]
\centering
\includegraphics[angle=0,width=0.43\linewidth]{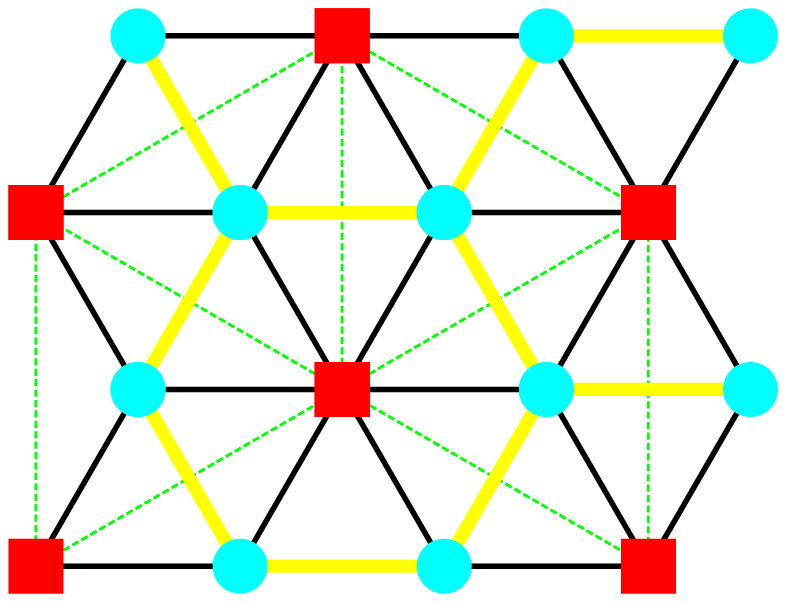} %
\includegraphics[angle=0,width=0.43\linewidth]{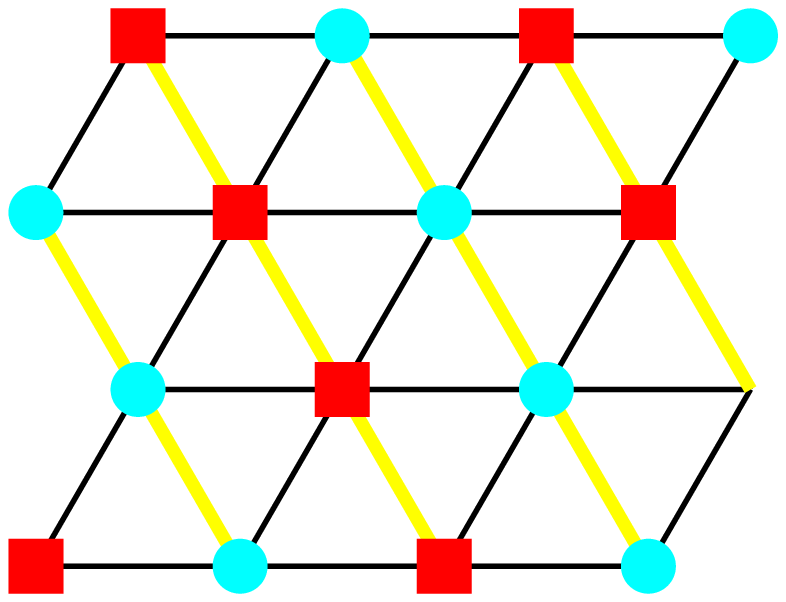}\newline
\includegraphics[angle=0,width=0.43\linewidth]{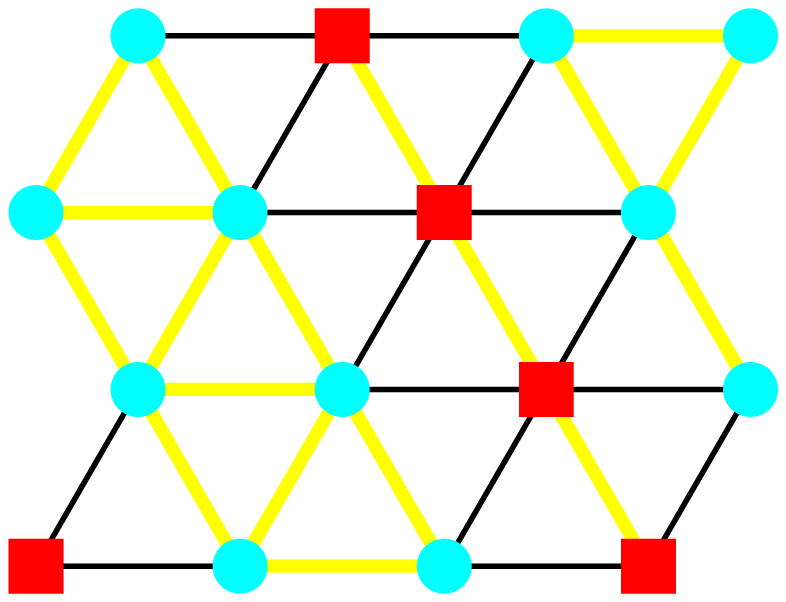} %
\includegraphics[angle=0,width=0.43\linewidth]{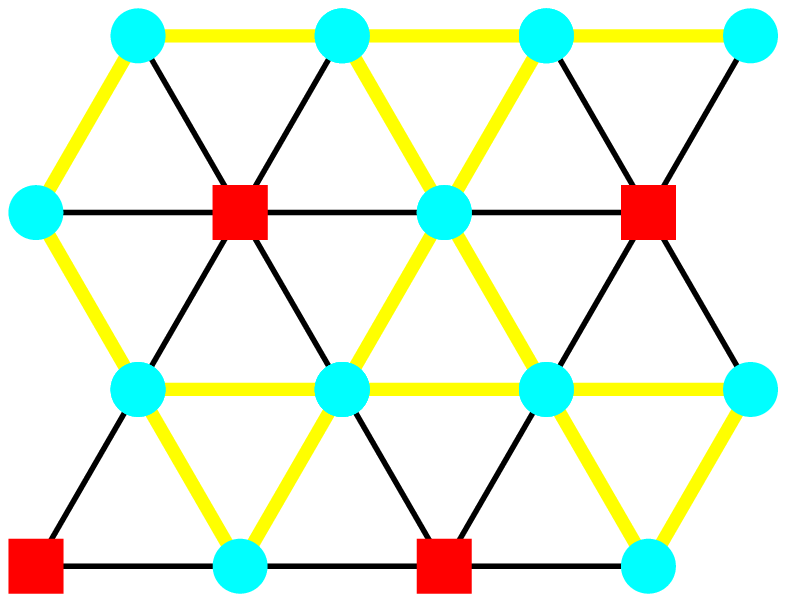}
\caption{(color online) Different magnetic patterns used in calculating the exchange
constants. Squares (circles) indicate up (down) moments within in the
supercell, thin dark (thick light) line antiferro- (ferro-)magnetic bonds.
The first pattern corresponds to the $\protect\sqrt{3}\times \protect\sqrt{3}
$ supercell indicated by the dash lines, the second, third and fourth
patterns to a $2\times 1$, $3\times 1$, and $2\times 2$ supercells,
respectively}
\label{str}
\end{figure}

Let us now investigate numerically the magnetic interactions in NiGa$_{2}$S$%
_{4}.$ First, we want to make sure that our conjecture about the absence of
magnetic anisotropy is indeed correct. This can be adressed by running fully
relativistic calculations imposing different magnetic field directions and
comparing energies. The result is that both energies differ by at most 0.03
meV. That is to say, the single site anisotropy is not an important factor
in NiGa$_{2}$S$_{4}.$ Having established that, we have computed several
different collinear magnetic structures, as shown in Fig. 2. If mapped onto
a Heisenberg model with three nearest neighbor interactions, these give the
exchange constants of 8.4, 0.3 and 4.1 meV (defined so that the total energy is
equal to sum over all bands of $J_{ij}S_{i}S_{j,}$ $|S_{i}|=1)$ for the first, 
second and third neighbors, respectively.

\begin{figure}[tbph]
\centering
\includegraphics[angle=0,width=0.93\linewidth]{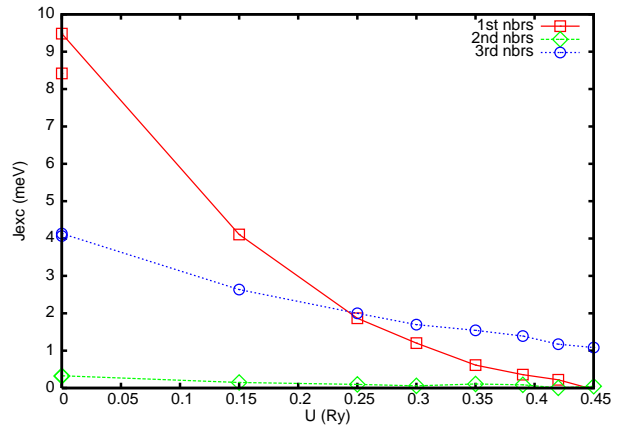}
\caption{(color online)
Calculated exchange constants  for the first three neighbor shells
in NiGa$_2$O$_4$, in meV, as a function of the Hubbard $U$, assuming 
the intraatomic $J=$0.07 Ry. The additional entry at $U=0$ corresponds to
$J=0$.}
\label{J}
\end{figure}

\begin{figure}[tbph]
\centering
\includegraphics[angle=0,width=0.93\linewidth]{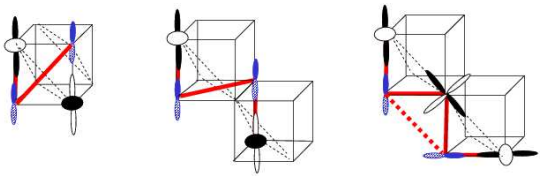}
\caption{(color online) Different possible superexchange paths within
 NiS$_2$ trilayers. Thin dashed triangles indicate the triangular Ni 
and (in the left panel) S layers. Thick (red online) lines show the exchange
paths. The solid parts correspond to short bonds (Ni-S, 2.422 \AA  and the shorter S-S bond, 
3.212 \AA), the dashed ones to the longer S-S bond, 3.626 \\AA.}
\label{hop}
\end{figure}

Several observations are in place. First, in LDA, while the second neighbor exchange
is negligible, the third one is sizeable and comparable with the nearest
neighbor exchange (it was suggested already in Ref. \onlinecite{Science} that the
3rd neighbor exchange may be anomalously large in this compound). This can
be traced down to anomalously large 3rd neighbor hopping, which is in fact
generic for triangular layers of transition metal oxides. Indeed, if the
metal-oxygen bonds form precisely 90$^{o}$ angles, the strongest nearest
neighbor hopping channel, $e_{g}-p-e_{g}$ (or, in compounds like Na$_{x}$CoO$%
_{2},$ $t_{2g}-p-t_{2g}$) is forbidden by symmetry, however, a third
neighbor path, $e_{g}-p-t_{2g}-p-e_{g}$ is fully allowed and in fact has the
most favorable geometry (Fig.\ref{hop})\cite{Miz}. This creates a possiblity for a sizeable
superexchange of the order of, as usually, $t_{eff}^{2}/\Delta ,$ where $%
t_{eff}$ is the effective hopping that appears after all intermediate states
are integrated out and $\Delta $ is the energy required to flip the spin of
a metal ion. In LDA, $t_{eff}$ is of the order of $t_{pd\sigma
}^{3}/(E_{d}-E_{p})/(E_{e_{g}}-E_{t_{2g}}),$ and $\Delta $ is of the order
of the Stoner (Hund) parameter, $\lesssim 1$ eV. In the Hubbard model, on
the other hand, $\Delta $ is set by the scale of Hubbard $U\sim 4-6$ eV. As
usually, the real life is somewhat in between, meaning that the exchange
constants in LDA are likely overestimated.

This can be easily demonstrated using the LDA+U method that takes into
account the Mott-Hubbard correlations in some crude approximation (Fig. \ref{J}).
To get an idea of the overall scale of the picture we have estimated 
the value of $U$ using the quasiatomic loop in a standard Linear Muffin Tin Orbital
package, as decribed in Ref. \onlinecite{Mol}, and obtained $U\approx 0.3 Ry$. 
This simplified method is known to underestimate $U$, therefore we have carried out
calculations up to $U=0.45$ Ry. At that maximal value of $U$ all three exchange
constants practically vanish, within the accuracy of the calculation. In fact,
the nearest neighbor constant at $U=0.45$ Ry becomes negative ($-0.04$ meV), but 
for all practical purpose it may be considered zero. Interestingly,
at this value of $U$ 
 the sum of our calculated
exchange constants over all (six) bonds gives 1.1*6=6.6 meV=76 K, to be 
compared with the Curie-Weiss temperature of 78$\pm 1$ K\cite{Science} (the calculated sign is 
antiferromagnetic, in agreement with the experiment).

What prevents the system from ordering remains unclear. 
Nakatsuji $et$ $al$\cite{Science} conjectured that the ratio 
of the nearest and the third neighbor exchange (no second neighbors)
is $\approx -0.2$, while our ratio, at large $U$, is essentially zero.
To this point it should be mentioned that 
neither the accuracy of LDA+U functional is sufficient to make firm statements
with a precision of a fraction of a meV, not the three-shell 
isotropic Heisenberg model is accurate to that extent. If the calculated numbers at $U=
0.45$ Ry are off by $\approx 0.2$ meV this would be enough 
to bring the calculated numbers in consistency with the 
Nakatsuji $et$ $al$'s\cite{Science} model.

Finally, one should keep in mind that the actual ordering temperature, if any, is
suppressed by the 2D character of magnetic interactions. Indeed, 
 NiGa$_2$O$_4$ has an extra GaO$_2$ layer compared to typical layer oxides
of the formula ABO$_2$ and therefore the surexchange interaction between the layers
is reduced. We have estimated this interaction by comparing the energies of the
fully ferromagnetic ordered structrure and the A-type antiferromagnetic one 
(FM layers stacked antiferromagnetically). The former is higher by about 1 meV
($J=$0.5 meV)
in LDA and by about 0.3 meV ($J=0.15$ meV) in LDA+U ($U=0.3$ Ry). Importantly, this superexchange
is additive with respect to all possible hopping paths between the layers that include
not only hopping from a Ni to the other Ni right above, but also to a large number
of neighboring Ni sites in the next plane\cite{Jp}. The real exchange coupling between the
two {\it antiferromagnetic} planes will be additionally reduced. An estimate
of $J_\perp\sim 0.05$ meV seems reasonable. Of course, although a computer code 
calculates the numbers with arbitrary precision, the actual physical approximations
used in the calculations preclude statements about energy differences of the order of
0.05 meV. For all practical purposes, it may be less than 0.01 meV, which, of course
would make long-range ordering at the experimentaly probed temperatures (a fraction of a Kelvin)
impossible.

To conclude, we have calculated the three nearest neighbor exchange constants 
in NiGa$_2$O$_4$ by mapping LDA+U calculations onto the isotropic Heisenberg
model. We found that (a) an anomalously large third neighbor coupling
exists in the system that can be traced down to superexchange $via$ occupied
$t_{2g}$ orbitals, (b) upon including correlation effects in terms of
Hubbard $U$, all exchange constants decrease rapidly, according to general
superexchange intuition, but the first neighbor exchange is more rapidly
suppressed than the third neighbor one, (c) at $U=0.45$
Ry (6 eV) the nearest neighbor exchange is entirely suppressed (and
we could not exclude, based on our calculations, that it does not become slightly
ferromagnetic), while the third one is exactly the right magnitude to explain
the observed Neel temperature. On the other hand, $U=6$ eV, at least on 
an intuitive level, seems to be too large for Ni in such an enviroment. At more realistic
$U$'s, such as 3-4 eV, both interactions remains firmly antiferromagnetic. 
It should be kept in mind that energy differences on the order of 1 meV are
on the borderline of many approximations used in our analysis. It is possible that several
weak effects conspire to prevent the system from ordering. First of all,
in particular the nearest neighbor interaction results from cancellation 
of two considerably stronger superexchange interactions of the opposite
signs: the AFM superexchange and the FM superexchange due to the Hund
rule coupling on S.
Albeit we see no obvious reason for the DFT to underestimate the latter,
such a possibility cannot be excluded.
Second, we did not make any attempt to evaluate further exchange interactions
beyond the third shell. While there is no special mechanism making them sizeable 
(as opposed to the third neighbor interaction), again we cannot prove 
that numerically.
Third, while the calculated on-site anisotropy is very
small, other effects beyond the isotropic Heisenberg model, such as 
biquadratic exchange, dipole-dipole, multispin interactions, etc\cite{Nagaev}
while small, may be not negligible on the backgound of the strong cancellation 
of the FM and the AFM superexchange, and may possibly create additional 
frustration in the system.  Finally, fourth, interplanar coupling between 
the antiferromagnetic noncollinearly correlated planes is at most a fraction of a
Kelvin, and possibly even smaller. This additionally supresses long range ordering.

The author is grateful to D.I. Khomskii for useful discussion and to C. Broholm
for bringing up this problem.

\end{document}